\input{aipcheck}

\documentclass[
    ,final            
  ]
  {aipproc}

\layoutstyle{6x9}

\begin{document}

\title{Accelerated expansion through interaction}


\author{Winfried Zimdahl}{
  address={Departamento de
F\'{\i}sica, Universidade Federal do Esp\'{\i}rito Santo \\
29075-910 Vit\'{o}ria, Esp\'{\i}rito Santo, Brazil}
}

\begin{abstract}
Interactions between dark matter and dark energy with a given
equation of state are known to modify the cosmic dynamics. On the
other hand, the strength of these interactions is subject to
strong observational constraints. Here we discuss a model in which
the transition from decelerated to accelerated expansion of the
Universe arises as a pure interaction phenomenon. Various
cosmological scenarios that describe a present stage
of accelerated expansion, like the $\Lambda$CDM model or a
(generalized) Chaplygin gas, follow as special cases for different
interaction rates. This unifying view on the homogeneous and
isotropic background level is accompanied by a non-adiabatic
perturbation dynamics which can be seen as a consequence of a
fluctuating interaction rate.
\end{abstract}

\maketitle


\section{Introduction}

 According to the prevailing view the dynamics of the present Universe is dominated by two so far unknown components, dark matter and dark energy. Dark matter is usually modeled as a pressureless fluid, while the preferred candidate for dark energy is a cosmological constant. The corresponding model, the $\Lambda$CDM model, is in good agreement with the observational data. However, there are competing approaches which also can provide a sufficiently large effective negative pressure, the key ingredient to account for a current stage of accelerated expansion within Einstein's General Relativity. Examples are scalar fields,
 a (generalized) Chaplygin gas, bulk viscous fluid models, holographic dark energy, to mention just a few (see, e.g., \cite{rev}).
 Most investigations in the field are
phenomenological and rely on the assumption that dark matter and dark energy evolve independently, i.e., their energies are assumed
to obey separate conservation laws. This implies
that the dark matter energy density varies as $a^{-3}$, where $a$
is the scale factor of the Robertson-Walker metric. The behavior
of the density of dark energy is then entirely governed by its
equation of state, the determination of which is a major subject
of current observational cosmology. However, there exists a line of research which takes into account the possibility of an interaction within the dark sector. Then, the evolution history of the Universe will be modified as the consequence of a coupling between dark matter and dark energy. In particular, the energy density of the (interacting) dark matter will no longer evolve as $a^{-3}$. There are also  modifications of the relation between luminosity distance and redshift. This class of interacting models typically starts with a given (in many cases constant) equation of state parameter for the dark energy and obtains interaction effects that are small corrections to the dynamics without interaction. The approach presented here is different. It uses arguments that grew out of the holographic principle. In this context an effective negative pressure is the result of an interaction in the dark sector. The transition from decelerated to accelerated expansion is understood as a pure interaction phenomenon.

\section{Interacting dark energy}

We assume the homogeneous and isotropic background dynamics to be given by
\begin{equation}
3\,\frac{\dot{a}^{2}}{a^{2}} \equiv 3\,H^{2} = 8\,\pi\,G\,\rho \ ,
\quad
\frac{\dot{H}}{H^{2}}\, = - \frac{3}{2}\left(1+
\frac{p}{\rho}\right)  \ \ \Rightarrow\ \ q \equiv -
\frac{\ddot{a}}{aH^{2}} = - 1 - \frac{\dot{H}}{H^{2}}
 \ ,\label{friedmann}
\end{equation}
where $\rho = \rho_{M} + \rho_{X}$ is the total energy density and $\rho_{M}$ and $\rho_{X}$ are the energy densities of
pressureless dark matter and dark energy, respectively.
The pressure of the $X$ component
coincides with the total pressure $p$. The quantity $H$ is the Hubble expansion rate, $q$ is the deceleration parameter and
$a$ is the scale factor of the Robertson-Walker metric.
We admit both components to be in interaction with each other according to
\begin{equation}
\dot{\rho}_{M} + 3H \rho_{M} = \frac{\dot{f}}{f}\rho_{M} \ ,\quad
\dot{\rho}_{X} + 3H (1+w)\rho_{X} = - \frac{\dot{f}}{f}\rho_{M}
 \ ,\label{balances}
\end{equation}
where the interaction is described by a time dependent function $f$.
While the overall energy density is conserved, the interaction induces a modification of the standard
$\rho_{M} \propto a^{-3}$ behavior of dark matter. Obviously, the interaction does not explicitly appear in \eqref{friedmann}. It needs the second derivative of the Hubble rate, equivalent to the third derivative of the scale factor, to see its influence directly \cite{statef}:
\begin{equation}
\frac{\ddot{H}}{H^{3}} = \frac{9}{2} +
\frac{9}{2}w\frac{\rho_X}{\rho}\left[2 + w +
\frac{1}{3H}\left(\frac{\dot{f}}{f}\frac{\rho_{M}}{\rho_{X}} -
\frac{\dot{w}}{w}\right)\right]
 \  .\label{}
\end{equation}
The impact on the dynamics is conveniently quantified by the ``statefinder" parameter
\begin{equation}
j\equiv \frac{1}{aH^3}\frac{\mbox{d}^3 a}{\mbox{d}t^3} = 1 +
3\frac{\dot H}{H^2} + \frac{\ddot H}{H^3}
 \  .\label{}
\end{equation}
This parameter enters the expression for the luminosity distance in third order,
\begin{equation}
d_L \approx \frac{z}{H_{0}} \left[1 + \frac{1}{2}\left(1 -q_0
\right)z + \frac{1}{6}\left( 3\left(q_0 + 1\right)^2 - 5\left(q_0
+ 1\right) + 1 - j_0 \right)z^2\right]
 \  .\label{}
\end{equation}
Here, the subscript $0$ denotes the present values of the corresponding quantities.
The influence of a non-vanishing interaction on the luminosity distance via $j_{0}$ should in principle be detectable, but is expected to be a small correction.

\section{Acceleration through interaction}
Defining the
ratio $r \equiv \frac{\rho_{M}}{\rho_{X}}$ of the energy densities of dark matter and dark energy, respectively, the dynamics of $r$ is determined by the balances \eqref{balances}. For the special case of a constant ratio we have
\begin{equation}
r = \mathrm{const} \quad \Leftrightarrow\quad w^{eff} = -
\frac{\dot{f}}{3 H f}\quad \Rightarrow\quad w = \left(1 + r
\right)w^{eff}\, .\ \label{rconst}
\end{equation}
Under this condition $w^{eff}$ is the \textit{total} equation of state of the
cosmic medium which also
coincides with the effective equations of state of the
components. Via Friedmann`s equation (\ref{friedmann}), a constant energy density ratio $r$ implies a dependence
$\rho_{X} \propto H^{2}$. Such a dependence is characteristic for holographic dark energy models in which the Hubble radius is chosen as the
infrared cutoff length \cite{cohen}.
We emphasize that
the effective equation of state parameter is entirely determined by the interaction. Without interaction one has $w^{eff} =  0$, the equation of state for dust, which is incompatible with an accelerated expansion of the Universe.
The corresponding deceleration parameter is \cite{WD}
\begin{equation}
q = \frac{1}{2}\left(1 - \frac{\dot{f}}{H f}\right)\
\ .\label{q}
\end{equation}
Consequently, we have decelerated expansion as long as
$\frac{\dot{f}}{f} < H$, while $\frac{\dot{f}}{f} > H$ is equivalent to accelerated expansion. A realistic description of our Universe requires a transition from
$\frac{\dot{f}}{f} < H$ to  $\frac{\dot{f}}{f} > H$.
Such a dynamics may be obtained if the
ratio $\frac{\dot{f}}{3 H f}$ is given in terms of a power of $H^{-1}$,
\begin{equation}
\frac{\dot{f}}{3 H f} = \mu
\left(\frac{H}{H_{0}}\right)^{-n} \qquad \Rightarrow\qquad
\dot{\rho} + 3 H\left(1 -
\mu\left(\frac{H}{H_{0}}\right)^{-n}\right)\rho = 0
\ ,\label{}
\end{equation}
where $\mu$ is an interaction constant \cite{essay}. For a spatially flat universe the
effective background equation of state then is proportional to $\rho^{-n/2}$.
This corresponds to a background energy density
\begin{equation}
\rho = \rho_{0}\left[\mu + \left(1 -
\mu\right)\left(\frac{a_{0}}{a}\right)^{3n/2}\right]^{2/n}
\ ,\label{rC}
\end{equation}
which is the energy density of a
generalized Chaplygin gas \cite{Bento}. In the special case $n = 2$ we recover the $\Lambda$CDM model in which now the interaction strength parameter $\mu$ determines
the cosmological constant.
We recall that we obtained (\ref{rC}) under the condition of a constant energy density ratio $r$. Nevertheless, a suitable matter dominated phase to ensure structure formation is guaranteed. The point here is that the interaction is negligible at early times for which the equation of state approaches that of non-relativistic matter. In other words, dark energy also behaves as matter at high redshifts.

\section{Perturbations}

Denoting first-order perturbations about the homogeneous and isotropic zeroth order by the hat symbol, we have
\begin{equation}
\hat{p} - \frac{\dot{p}}{\dot{\rho}}\hat{\rho} = p
\left[\frac{\hat{R}}{R} -
\frac{\dot{R}}{R}
\frac{\hat{\rho}}{\dot{\rho}}\right] = \frac{n}{2} p
\left(\frac{\hat{\rho}}{\rho} - 2 \frac{\hat{H}}{H}\right)
\ , \label{hp}
\end{equation}
where $R\equiv  \frac{\dot{f}}{Hf}$ is the ratio of the interaction rate to the Hubble rate and $\hat{R}$ is the perturbation of this ratio. In the general case, the right hand side of (\ref{hp})  is different from zero which means that the perturbations are non-adiabatic and do \textit{not} propagate with the adiabatic sound speed $\frac{\dot{p}}{\dot{\rho}}$.
One can show that the relation between the perturbations of pressure and energy density for a comoving observer is
``non-local" \cite{essay}. It is not just determined algebraically by a given equation of state parameter but becomes part of the dynamics. Effects of this type are relevant for the matter power spectrum.\cite{bv}.

\section{Discussion}

A coupling between dark matter and dark energy is commonly believed to result in small corrections to the uncoupled dynamics, subject to strong observational constraints.
Here we have established an approach in which interactions
in the dark sector of the cosmological substratum are the essential ingredient to obtain an effective negative pressure.
The transition from decelerated to accelerated expansion of the Universe may be seen as a pure interaction effect in the context of a holographic dark energy. In the homogeneous and isotropic background the $\Lambda$CDM  and the (generalized) Chaplygin gas models are recovered as subcases for different interaction rates. However, the corresponding perturbation dynamics is intrinsically non-adiabatic, i.e.,  there are differences to the ``standard" adiabatic $\Lambda$CDM  and  Chaplygin gas models which are expected to be observationally relevant.



\begin{theacknowledgments}
I am grateful to the organizers of the ERE2008 meeting in Salamanca, especially to
Fernando Atrio-Barandela. I acknowledge support by Brazilian grants 093/2007 (CNPq and FAPES) and EDITAL
FAPES No. 001/2007.
\end{theacknowledgments}





\begin{thebibliography}{9}

\bibitem{rev}
E.~J.~Copeland, M.~Sami and S.~Tsujikawa, \emph{Int. J. Mod. Phys. D} \textbf{15}, 1753
(2006);
R.~Durrer and R.~Maartens, \emph{Gen. Rel. Grav.} \textbf{40}, 301
(2008).

\bibitem{statef} W.~Zimdahl and D.~Pav\'{o}n, \emph{Gen. Rel. Grav.} \textbf{36}, 1483 (2004).

\bibitem{cohen}
A.~G.~Cohen, D.~B.~Kaplan and A.~E.~Nelson, \emph{Phys. Rev. Lett.}
\textbf{82}, 4971 (1999).

\bibitem{WD}
W.~Zimdahl and D.~Pav\'{o}n, \emph{Class. Quantum Grav.} \textbf{24}, 5641
(2007).

\bibitem{essay}
W.~Zimdahl, \emph{Int. J. Mod. Phys. D} \textbf{17}, 651
(2008).

\bibitem{Bento}
M.~C.~Bento, O.~Bertolami and A.~A.~Sen, \emph{Phys. Rev. D} \textbf{66},
043507 (2002).

\bibitem{bv} R.~Colistete~Jr., J.~C.~Fabris, J.~Tossa and W.~Zimdahl,
\emph{Phys. Rev. D} \textbf{76}, 103516 (2007).




\end{thebibliography}

%

\end{document}